\documentclass[conference]{IEEEtran}
\IEEEoverridecommandlockouts

\usepackage{amsmath,amssymb,amsfonts,accents,mathrsfs}
\usepackage{mathtools}
\usepackage{array}

\usepackage{float}
\usepackage{bm}

\usepackage{graphicx}
\usepackage{textcomp}
\usepackage[svgnames]{xcolor} 
\usepackage{subfloat}

\definecolor{amethyst}{rgb}{0.6, 0.4, 0.8}

\usepackage[labelsep=quad,indention=10pt]{subfig}
\usepackage{algorithm}
\usepackage{algorithmic}

\usepackage{tikz}
\usetikzlibrary{positioning}
\usepackage{textcomp}
\usetikzlibrary{shapes,arrows,backgrounds}
\usetikzlibrary{datavisualization}
\usetikzlibrary{patterns}

\definecolor{x11_gray}{rgb}{0.85, 0.85, 0.85}

\def\BibTeX{{\rm B\kern-.05em{\sc i\kern-.025em b}\kern-.08em
    T\kern-.1667em\lower.7ex\hbox{E}\kern-.125emX}}

\renewcommand{\t}{\left[t\right]}

\usepackage{pgfplots}
\usepackage{tikz}
\usepgflibrary{plotmarks}
\usepgflibrary[plotmarks]
\usetikzlibrary{plotmarks}
\usetikzlibrary[plotmarks]
\usetikzlibrary{positioning}
\usepackage{textcomp}
\usetikzlibrary{shapes,arrows,backgrounds}
\usetikzlibrary{datavisualization}
\usetikzlibrary{patterns}
\pgfplotsset{compat=newest}
\pgfplotsset{plot coordinates/math parser=false}
\newlength\figureheight
\newlength\figurewidth

\usepackage{mathtools}

\newcommand{\tabcomplex}{
    \begin{tabular}{lc} \hline
        \footnotesize Algorithm          & \footnotesize Complexity \\ \hline
        \footnotesize Linear MMSE        & $\scriptstyle O(N^3)$ \\
        \footnotesize SA-SIC \cite{Knoop2013}     & $\scriptstyle O(N^3)$ \\
        \footnotesize AA-MF-SIC \cite{DiRenna2019} & $\scriptstyle O(N^3)$ \\
        \footnotesize AA-RLS \cite{JChoi2005}        & $\scriptstyle O(N^2)$ \\
        \footnotesize AA-RLS-DF             & $\scriptstyle O(N^2)$ \\
        \footnotesize AA-CL-RLS             & $\scriptstyle O(N^2)$ \\
        \footnotesize AA-CL-DF & $\scriptstyle O(N^2)$ \\ \hline
    \end{tabular}}

\newcommand{\tabvar}{
    \begin{tabular}{lc} \hline
        \footnotesize Parameter        & \footnotesize Value \\ \hline
        \footnotesize Num. devices $N$ & $128$ \\
        \footnotesize Spread. gain $M$ & $64$ \\
        \footnotesize Activity prob.   & $[0.1,0.3]$ \\
        \footnotesize Modulation       & QPSK \\
        \footnotesize Block size       & $128$ symb. \\
        \footnotesize Metadata         & $60$ \\
        \footnotesize Data             & $68$ \\ \hline
    \end{tabular}}

\begin{document}

\title{Study of Adaptive Activity-Aware Constellation List-Based Detection for Massive Machine-Type Communications\\
\thanks{This work was supported in part by the National Council for Scientific and Technological Development (CNPq) and in part by FAPERJ.}
}

\author{\IEEEauthorblockN{Roberto B. Di Renna, \textit{Student Member}, \textit{IEEE} and Rodrigo C. de Lamare, \textit{Senior Member}, \textit{IEEE}}
\IEEEauthorblockA{Center for Telecommunications Studies (CETUC), Pontifical Catholic University of Rio de Janeiro, RJ, Brazil}}

\maketitle

\begin{abstract}
In this work, we propose an adaptive list-based decision feedback detector along with an $l_0$-norm regularized recursive least-squares algorithm that only requires pilot symbols (AA-CL-DF). The proposed detector employs a list strategy based on the signal constellation points to generate different candidates for detection. Simulation results show that the proposed AA-CL-DF successfully mitigates the error propagation and approaches the performance for the oracle LMMSE algorithm.
\end{abstract}


\section{Introduction}

Massive machine type communications (mMTC) is one of the main features of the fifth generation (5G) of wireless communications systems. With its uplink-focused traffic, a large number of devices send small packets of up to a few bits sporadically~\cite{Chen2017}, a scenario that differs from well-known multiple-antenna systems.

Prior work on channel estimation and detection of devices includes compressed sensing (CS) methods that outperform conventional channel estimation and detection techniques~\cite{JChoi2017}. Nonetheless, if the device activity detection is accurate, then conventional techniques may be employed. As mMTC detection is an open problem, most of the works to date investigate different approaches~\cite{Zhu2011,Knoop2013,Ahn2018,DiRenna2019}.

In this paper, we introduce a novel strategy to improve a decision feedback detector along with an activity-aware $l_0$-norm regularized recursive least squares (RLS) algorithm (AA-CL-DF). Inspired by the concept of lists of vector candidates, we mitigate the error propagation with the aid of a successive interference cancellation (SIC) scheme and list of symbol vector candidates. Simulations show that the AA-CL-DF outperforms prior work with a competitive complexity.

The rest of this work is organized as follows: Section II describes the system model. Section III presents the proposed adaptive activity-aware constellation list decision feedback detection, whereas Section IV details the proposed soft information processing and decoding scheme. Section V examines the complexity analysis of the proposed and existing techniques, while the simulation results are shown and discussed in Section VI. Section VII gives the conclusions of the work.

\section{System model}
\label{sec:System_Model}

\begin{figure}[t]
\subfloat[Uplink
scenario.\label{fig:mtc_system}]{\includegraphics[scale=.4]{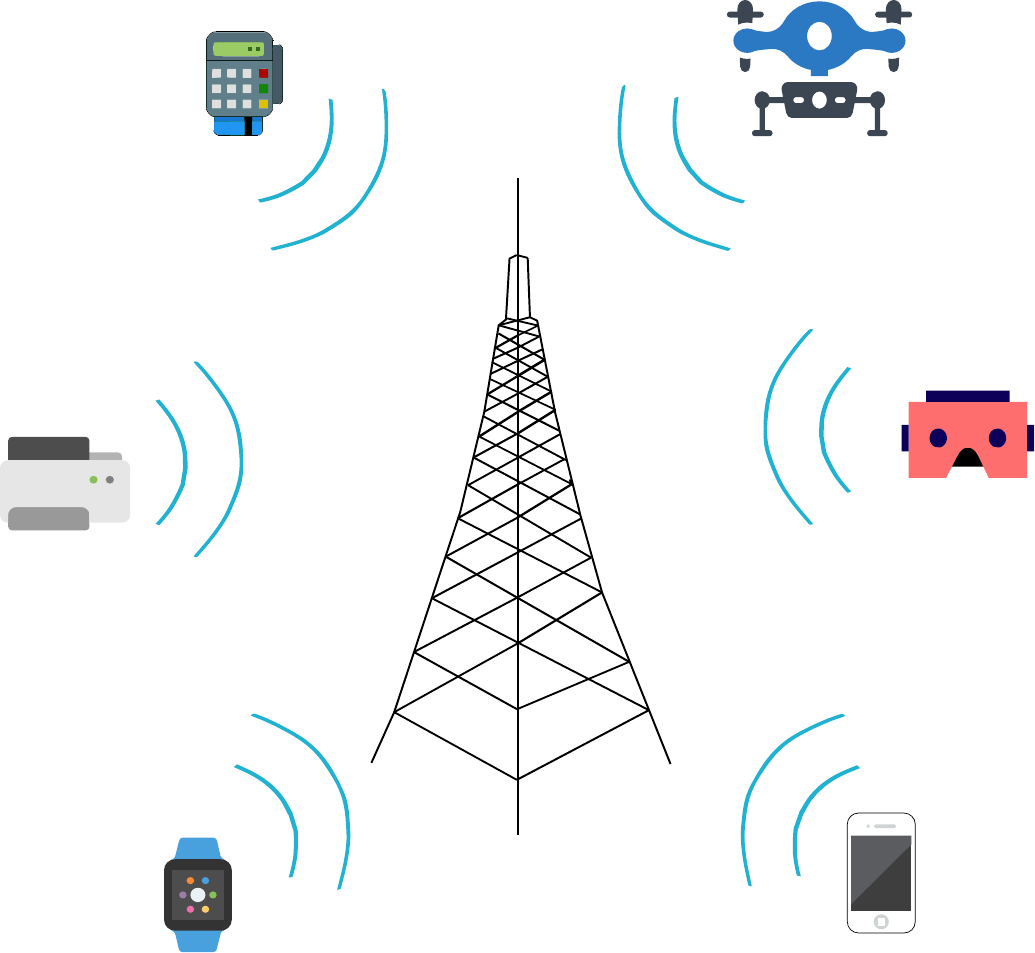}}\hfill
\subfloat[Reliability radius for $\mathcal{A}_0$
alphabet.\label{fig:qpsk_aug}]{\includegraphics[scale=0.8]{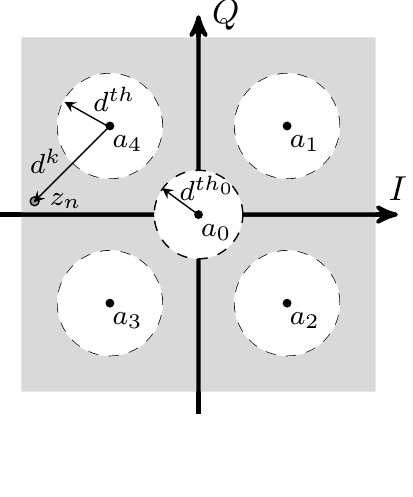}}
\caption{(a) mMTC uplink multiple access scenario with N devices
sporadically active and (b) augmented QPSK alphabet.}
\label{fig:conj}
\end{figure}

In this work we assume the transmitted data symbols are distributed
as zero-mean Gaussian sequences, in a similar way to Low-Active Code
Division Multiple Access (LA-CDMA) scenario~\cite{deLamare2008}.
Alternatively, one can also consider an access point equipped with
an antenna array with $M$ antenna elements 
\cite{Vantrees1,locsme,elnashar,manikas,cgbf,okspme,r19,scharf,bar-ness,pados99,
reed98,hua,goldstein,santos,qian,delamarespl07,delamaretsp,xutsa,xu&liu,
kwak,delamareccm,delamareelb,wcccm,delamarecl,delamaresp,delamaretvt,delamaretvt10,delamaretvt2011ST,
delamare_ccmmswf,jidf_echo,jidf,barc,lei09,delamare10,fa10,ccmavf,lei10,jio_ccm,
ccmavf,stap_jio,zhaocheng,zhaocheng2,arh_eusipco,arh_taes,rdrab,dcg,dce,dta_ls,
song,wljio,barc,saalt,mmimo,wence,spa,mbdf,rrmber,bfidd,did,mbthp,wlbd,baplnc,lrcc}.
In the multi-user uplink context, $N$ devices transmit data and
metadata (pilots) taken from the constellation set $\mathcal{A}$ at
each time instant to a central aggregation node. Considering a
different per device activity probability $p_n$, each symbol $x_n$
is drawn from the augmented alphabet $\mathcal{A}_0$, structured as
an equi-probable finite alphabet $\mathcal{A}$ (QPSK, for example),
when the $n$-th device is active and zero otherwise ($\mathcal{A}_0
= \mathcal{A}\cup 0$). The received signal associated with a
spreading factor $M$ is given by:

\begin{equation}\label{eq:main}
    \mathbf{y} = \mathbf{H}\mathbf{x} + \mathbf{v},
\end{equation}
\noindent where the matrix $\mathbf{H} \in \mathbb{C}^{M\times N}$
contains the spreading sequences and channel impulse responses to
the base station (BS), while $\mathbf{v}$ is the additive white
Gaussian noise with zero mean and covariance matrix
$\mathbb{E}\left[\mathbf{v}\mathbf{v}^\text{H}\right]=\sigma_v^2
\mathbf{I}$. Fig.~\ref{fig:mtc_system} represents the scenario.

\section{Proposed Adaptive Activity-Aware Constellation List Decision Feedback Detection}
\label{sec:AA-CL-DF}

In this section, we detail the proposed adaptive activity-aware constellation list decision feeedback detecion scheme that is inspired by prior work on decision feedback techniques \cite{deLamare2008,mbdf} and list-based strategies \cite{Fa2011,DiRenna2019}.

\subsection{Adaptive Detection Design} \label{subsec:adapt_imp}

Considering $i$ as the time instant, the feedforward filters are updated at each iteration and the feedback ones cancel the already-determined signals~\cite{JChoi2005}. In this way, the vector which refers to feedforward and feedback filters can be written as

    \begin{equation}\label{eq:filter_forw_fedd}
        \mathbf{w}_{n}\left[i\right] =
        \left\{
        \begin{array}{cl}
            \mathbf{w}^{f}_{n}\left[i\right], & n=1; \\
            \left[{\mathbf{w}^{f}_{n}}^\text{T}\left[i\right],{\mathbf{w}^{b}_{n}}^\text{T}\left[i\right]\right]^\text{T}, &    n=2,\dots, N.
        \end{array}\right.
    \end{equation}

\noindent and the augmented received vector,

    \begin{equation}\label{eq:all_y}
        \mathbf{y}_{n}\left[i\right] =
        \left\{
        \begin{array}{cl}
            \mathbf{y}\left[i\right], & n=1; \\
            \left[{\mathbf{y}}^\text{T}\left[i\right],{\hat{\mathbf{d}}^\text{T}_{\phi_{n-1}}}\left[i\right]\right]^\text{T}, & n=2,\dots, N.
        \end{array}\right.
    \end{equation}

\noindent where $\mathbf{w}_{n}\left[i\right] = \left[w^{f}_{n,1}\left[i\right], \dots, w^{f}_{n,M}\left[i\right], w^{b}_{n,M+1}\left[i\right],\dots,\right.$\\ $\left. w^{b}_{n,M+N}\left[i\right]\right]^\text{T}$ corresponds of both filters for the detected symbol of the $n$-th device. The vectors $\hat{d}_{\phi_n}\left[i\right]$ and $\tilde{d}_{\phi_n}\left[i\right]$ denote the $n$-th detected symbols and corresponding output, respectively, where $\phi_n \in \left\{1,2,\dots,N\right\}$. The output of the detectors can then be represented as

    \begin{equation}\label{eq:out_equa}
        \tilde{d}_{\phi_n}\left[i\right] = \mathbf{w}_n^\text{H}\left[i\right]\mathbf{y}_n\left[i\right].
    \end{equation}

The receive filter is computed by the regularized RLS algorithm, whose cost function is given by

    \begin{eqnarray}\label{eq:cost_func_rls}
    \mathcal{J}_{n,j}\left[i\right]\hspace{-8pt} &=&\hspace{-10pt}  \sum_{l=0}^{i} \lambda^{i-l} \left|\hat{d}_{j}\left[l\right]-\mathbf{w}_n^\text{H}\left[i\right]\mathbf{y}_n\left[i\right]\right|^2 + \, \gamma \|\mathbf{w}_{n}\left[i\right]\|_0
    \end{eqnarray}

\noindent where $0< \lambda \leq 1$ is the forgetting factor and $\|\cdot\|_0$ denotes $l_0$-norm that counts the number of zero entries in $\mathbf{w}_{n}$ and $\gamma$ is a non-zero positive constant to balance the regularization and, consequently, the estimation error.

Approximating the value of the $l_0$-norm~\cite{Bradley98}, the regularization in (\ref{eq:cost_func_rls}) can be rewritten as

    \begin{equation}\label{eq:cost_func_rls_l0_app}
         \|\mathbf{w}_{n}\left[i\right]\|_0 =  \sum^{2M}_{p=1} \left(1-\text{exp}\left(-\beta|w_{n,p}\left[i\right]|\right)\right).
    \end{equation}

The parameter $\beta$ regulates the range of the attraction to zero on small coefficients of the filter~\cite{Bradley98}. Thus, taking the partial derivatives for all entries $i$ of the coefficient vector $\mathbf{w}^{t}_{n}$ in (\ref{eq:cost_func_rls_l0_app}), setting the results to zero  and after some approximations, we obtain

    \begin{eqnarray}\nonumber
        \mathbf{w}_{n}\left[i\right] &=& \mathbf{w}_{n}\left[i-1\right] + \mathbf{k}\left[i\right] \epsilon_n^\ast\left[i\right] \\ \label{eq:recursive_rls_l0_app}
        &&  - \gamma \, \beta\, \text{sgn}\left(w_{n,p}\left[i\right]\right)f_\beta\left(w_{n,p}\left[i\right]\right)
    \end{eqnarray}

\noindent where the function $f_\beta\left(w_{n,p}\left[i\right]\right)$ is given by

    \begin{equation}\label{eq:f_beta}
        f_\beta\left(w_{n,p}\left[i\right]\right) =
        \left\{
        \begin{array}{rl}
            \beta^2 \left(w_{n,p}\left[i\right]\right)+\beta, & -1/\beta \leq w
            _{n,p}\left[i\right]< 0; \\
            \beta^2 \left(w_{n,p}\left[i\right]\right)-\beta, & 0< \leq w_{n,p}\left[i\right]\leq 1/\beta; \\
            0,                           & \text{otherwise.}
        \end{array}\right.
    \end{equation}

To determine the detection order $\phi_n$ for the $n$-th DF detector, the set of candidate symbols are evaluated. Considering $S_i$ as all indices of the symbols to be detected from the $n$-th iteration ($S_i = \left\{1,2,\dots,N\right\} - \left\{\phi_1,\dots,\phi_{n-1}\right\}$), the detection order is determined by selecting the symbol associated with the minimum LSE, as given by

    \begin{equation}\label{eq:detec_ord}
        \phi_n = \underset{j\in S_i}{\textrm{arg min}}\hspace*{5pt} \mathcal{J}_{n,j}\left[i\right].
    \end{equation}

As both the detection order and weights are updated at each time $i$, we take a decision using the weight and the received vector $\mathbf{y}$ at time $i-1$, respecting the detection order $\phi_n$.

\subsection{Constellation-type lists}

The constellation-type list is based on the multi-feedback SIC \cite{Fa2011} and the AA-MF-SIC method, first presented in~\cite{DiRenna2019}, that verifies if the previous estimate is reliable or not. To judge that, a shadow area constraint is proposed, as in Fig.\ref{fig:qpsk_aug}. If the symbol drops into the shadow area, it is considered unreliable, otherwise it is considered reliable.

Due to the fact that the augmented QPSK symbols ($a_f$) do not have the same \textit{a priori} probabilities, there are different radius of reliability. As shown in~\cite{DiRenna2019}, better performance is achieved if for the zero, the radius of reliability considered is $\left(1- 1/\lambda_n\right)$ and, otherwise, $d^{th} = 1/\lambda_n$. Then, the presence of the SAC reduces computational complexity.

If the soft estimate, after the SAC, is considered reliable, a hard slice will be performed and the symbol $\hat{x}_{n}$ will be used in the feedback for the next stage. Otherwise, a candidate vector called CL set $\mathcal{L}=\left[c_1,c_2,\dots,c_k,\dots,c_K\right] \subseteq \mathcal{A}_0$ is generated, where is compound by the $K$ nearest constellation points to the soft decision $z_{n}$. The candidate vector in the $n$-th stage, is given by

    \begin{align}
        \mathbf{b}^k =& \left[\underbrace{z_{1},\cdots,z_{n-1}}_{\text{already detected symbols}},c_k,\cdots,\underbrace{b_{j}^k}_{\mathcal{Q}\left[\mathbf{w}^\text{H}_{j}\mathbf{y}^k_{j}\right]},\cdots,b_{N}^k\right]^\text{T}
    \end{align}

\noindent where the candidate $c_k$ is used as the cancellation symbol in the $n$-th  stage and successively processing the remaining $n+1$ to $N$ stages with the conventional SIC as

    \begin{align}
        \mathbf{y}_{j}^k \hspace{10pt} &= \hspace{10pt} \mathbf{y}_{n}^k - \overline{\mathbf{H}}_n\, c_k - \sum_{p=n+1}^{j-1} \mathbf{H}_p b_{p}^k.
    \end{align}

\noindent where $\overline{\mathbf{H}}_{n}$ denotes the matrix obtained by taking the columns $n, n+1, \dots , N$ of $\overline{\mathbf{H}}$ . Using the ML rule, the list selects the best candidate according to

\begin{equation}
    k_\text{opt} = \underset{1 \leq k \leq K}{\textrm{arg min}}\hspace*{5pt} \|\mathbf{y} - \mathbf{H}\,\mathbf{b}^k\|^2.
\end{equation}

\begin{figure}[t]
\centering
    \includegraphics[scale=1]{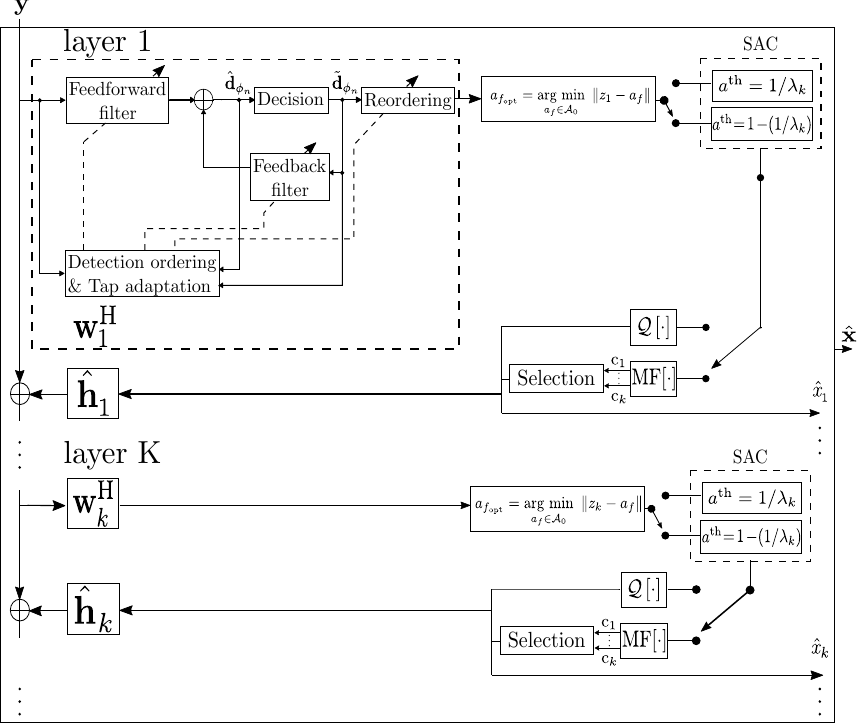}
    \caption{Block diagram of the AA-CL-DF scheme.}
    \label{fig:diagram}
\end{figure}

The $c_{k_\text{opt}}$ is chosen to be the optimal feedback symbol for the next stage as well as a more reliable decision for the current stage. Fig.~\ref{fig:diagram} shows the block diagram that describes the proposed AA-CL-DF algorithm.

\section{Proposed Soft Information Processing and Decoding} \label{sec:prop_it_det_dec}

We draw inspiration on~\cite{XWang1999} and \cite{bfidd} to propose an IDD scheme that estimates and incorporates the probability of each device being active in the mMTC scenario. In order to reduce the complexity of the system, as our scheme does not require matrix inversions, the complexity of the algorithm is lower than other approaches.

As each device has a different activity probability $\rho_n$, the \textit{a priori} probabilities $\text{Pr}\left(x_n\t = \overline{x}\right)$ should take into account, as

\begin{align}
    \begin{array}{rl}
     \hspace{-4pt}\rho_n + \left(1-\rho_n\right)\text{Pr}\left(x_n\t = \overline{x}\right), & \hspace{-7pt} {\text{if}\, \left(\overline{x}^1 \text{ and } \overline{x}^2\right) = 0,}    \\ \label{eq:a_priori2}
      \left(1-\rho_n\right)\text{Pr}\left(x_n\t = \overline{x}\right), & \hspace{-7pt} \text{otherwise}
    \end{array}
\end{align}

\noindent where the \textit{a priori} probabilities are computed based on the extrinsic LLRs $L_{e_n}^z\t$, provided by the LDPC decoder, as

    \begin{equation}\label{eq:a_priori1}
        \text{Pr}\left(x_n\t = \overline{x}\right) = \sum_{\overline{x}\in \mathcal{A}_0} \overline{x}\left(\prod^{M_c}_{z=1}\left[1+\text{exp}\left(-\overline{x}^z L_{e_n}^z\t\right)\right]^{-1}\right),
    \end{equation}
\noindent where $M_c$ represents the total number of bits of symbol $\overline{x}$, the superscript $z$ indicates the $z$-th bit of symbol of $\overline{x}$, in $\overline{x}^z$ (whose value is $(+1,-1)$). The bits are assumed to be statistically independent of one another.

In order to initialize the iterative scheme, all $L_{e_n}^z\t$ are zero and, in the next iteration of the scheme, the new \textit{a priori} probabilities incorporates the probability that the $n$-th device is active and the updated extrinsic LLR values.

As the output of the proposed receive filter has a large number of independent variables, we can approximate it as a Gaussian distribution~\cite{XWang1999}. Hence, we approximate $\hat{d}_{\phi_n}\t$ by the output of an equivalent AWGN channel with $\hat{d}_{\phi_n}\t = \mu_n\left[i\right] x_n\t + b_n\t$. Where at first, $\mu_n\t = \mathbf{w}_n^\text{H}\t \mathbf{y}_n\t$, while for the following cases all $x_n$ are the previously detected symbols. Thereat, the likelihood function $P\left(\hat{d}_{\phi_n}|\overline{x}\right)$ is approximated by

    \begin{equation}\label{eq:likelihood}
        P\left(\hat{d}_{\phi_n}\t|\, x\right) \approx \frac{1}{\pi\, \zeta^2_n\t} \text{exp}\left(-\frac{1}{\zeta_n^2\t}|\hat{d}_{\phi_n}\t-\mu_n\t \overline{x}|^2\right),
    \end{equation}

\noindent where the mean $\mu_n\t = \mathbb{E}\left\{\hat{d}_{\phi_n}\t x_{n}\t\right\}$ can be approximated by

    \begin{eqnarray}\label{eq:mean_gauss_app}
        \mu_n\t &\approx & \mathbf{w}_n^\text{H}\t \left(\sum_{p=1}^{t-1}\lambda^{t-1-p}\, \mathbf{y}_n\left[p\right]x_{n}\left[p\right]\right).
    \end{eqnarray}

As long as $b_n\t$ are zero-mean complex Gaussian variables with variance $\zeta^2_n\t = \text{var}\left\{\hat{d}_{\phi_n}\t\right\}$, we can write
    \begin{eqnarray}\nonumber
        \zeta_n^2\t &\approx& \mathbf{w}_n^\text{H}\t\left(\sum_{p=1}^{t}\lambda^{t-p}\, \mathbf{y}_n\left[p\right]\mathbf{y}_n^\text{H}\left[p\right]\right)\mathbf{w}_n\t - \mu_n^2\left[i\right]. \\ \label{eq:var_gauss_app}
    \end{eqnarray}

Then, the extrinsic LLRs computed by the AA-VGL-DF detector for the $z$-th bit ($z \in \{1,\dots,M_c\}$) of the symbol $x_n$ transmitted by the $n$-th device are given by

\begin{align}\label{eq:llr}
        L_{c_n}^z\t =& \log \frac{\sum_{\overline{x}\in \mathcal{A}_z^{+1}}\text{Pr}\left(\hat{d}_{\phi_n}\t|\, \overline{x}\right)\text{Pr}\left(\overline{x}\right)}{\sum_{\overline{x}\in \mathcal{A}_z^{-1}}\text{Pr}\left(\hat{d}_{\phi_n}\t|\, \overline{x}\right)\text{Pr}\left(x\right)} - L_{e_n}^z\t
\raisetag{20pt}
\end{align}

\noindent where $\mathcal{A}_z^{+1}$ is the set of $2^{M_c -1}$ hypotheses of $\overline{x}$ for which the $z$-th bit is +1 (analogously for $\mathcal{A}_z^{-1}$).

\begin{table}[t]
    \begin{center}
        \begin{tabular}{p{0.2cm}l} \\ \hline
        \multicolumn{2}{l}{\textbf{Algorithm 1} Proposed IDD with AA-CL-DF} \\ \hline
            1. & Initialization: $M$, $N$, $\bm{\rho}$, $\xi$, $\gamma$, $\lambda$, $\mathbf{P}_n = \bm{\rho}\,\mathbf{I}_{\text{M}}$\\
               & \footnotesize\textbf{ \% For training mode,} \\
               & \footnotesize \hspace{10pt} \% For each metadata sequence $\hat{\mathbf{d}}_p\t$ and $\mathbf{y}_{p,n}\t$, \\
            \small
            2. & \hspace{10pt} Compute the Kalman gain vector\\
               & \hspace{10pt} $\mathbf{k}_n\t = (\mathbf{P}_n\t \,\mathbf{y}_{p,n}\t)/(\lambda+\mathbf{y}_{p,n}^\text{H}\t\, \mathbf{P}_n\t \, \mathbf{y}_{p,n}\t)$;\\
            3. & \hspace{10pt} Estimate $\tilde{d}_{\phi_n}\t = \mathbf{w}_n^\text{H}\t\,\mathbf{y}_{p,n}\t$;\\
            4. & \hspace{10pt} Update the error value with $\epsilon_{\phi_n}\t = \hat{d}_{p,\phi_n}\t - \tilde{d}_{\phi_n}\t$;\\
            5. & \hspace{10pt} Update the filters with Eq.~(\ref{eq:recursive_rls_l0_app});\\
            6. & \hspace{10pt} Update the auxiliary matrix\\
               & \hspace{10pt} $\mathbf{P}_n\t = \lambda^{-1} \left(\mathbf{P}_n\t - \mathbf{k}_n\t\, \mathbf{y}_{p,n}^\text{H}\t\,\mathbf{P}_n\t\right)$; \\
            7. & \hspace{10pt} Concatenate $\mathbf{y}_n\t$ with $\hat{d}_{\phi_{p,n}}\t$; \\
            8. & \hspace{10pt} Update the sequence of detection with Eq.(\ref{eq:detec_ord}); \\
               & \footnotesize \textbf{\% For decision-directed mode,}\\
               \small
            9. & \hspace{10pt} Compute the \textit{a priori probability} with Eqs.~(\ref{eq:a_priori1}) and (\ref{eq:a_priori2}); \\
           10. & \hspace{10pt} Repeat steps $2.$ to $6.$; \\
           11. & \footnotesize \hspace{10pt} Evaluate the reliability of the soft estimation $\tilde{d}_{\phi_n}\t$ in the SAC \\
               & \footnotesize \hspace{10pt} and proceeds with the internal list if it is judged as unreliable; \\
           12. & \hspace{10pt} Update the sequence of detection with the output of 11; \\
           \small
           13. & \hspace{10pt} Compute $\mu_{\phi_n}\t$ and $\zeta^2_{\phi_n}\t$ with Eqs.~(\ref{eq:mean_gauss_app}) and (\ref{eq:var_gauss_app});\\
           14.  & \hspace{10pt} Verify the likelihood function $P\left(\hat{d}_{\phi_n}\t|x\right)$ with Eq.~(\ref{eq:likelihood}); \\
           15.  & \hspace{10pt} Compute the LLR value according to Eq.(\ref{eq:llr}). \\ \hline
        \end{tabular}
    \end{center}
\end{table}

\section{Complexity analysis}
\label{sec:compl}

The computational complexity of the proposed AA-CL-DF scheme is given this section. In a nutshell, the complexity of AA-CL-DF is given by the regularized RLS complexity added by that of the CL algorithm. Considering $N=M$ and $K$ as the number of candidates in the CL, AA-CL-DF has the same order complexity as the state-of-the-art of algorithms, as seen in Table~\ref{tab:complexity}.

\begin{figure}[t]
    \centering
    \includegraphics[scale=.825]{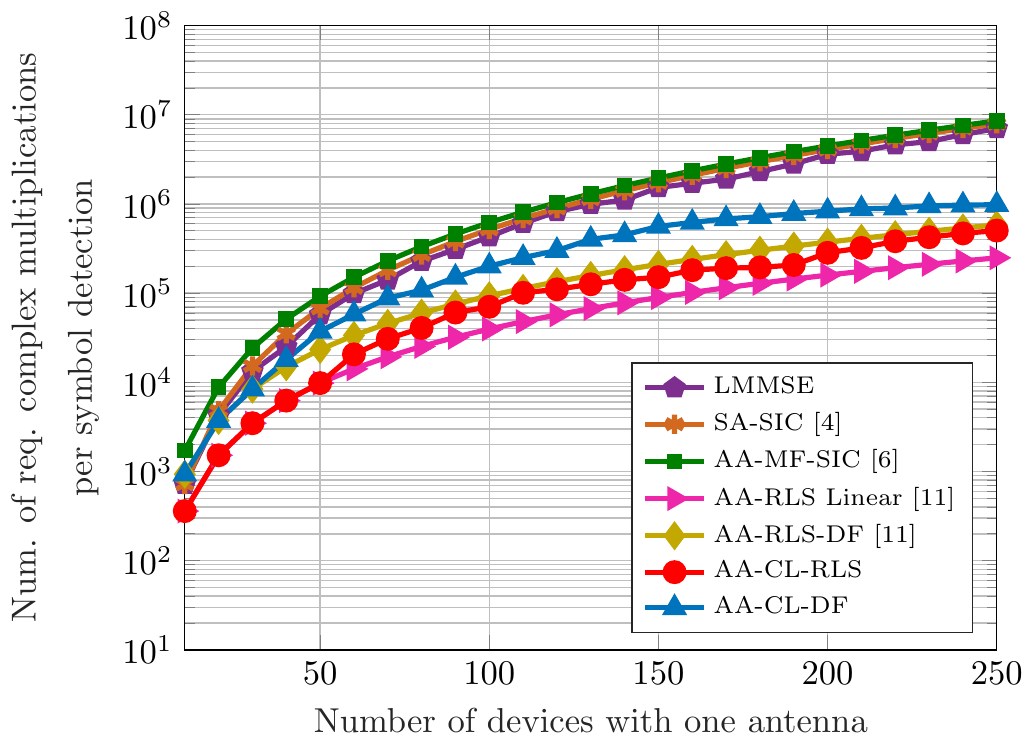}
    \caption{Comparison of complexity of considered algorithms.}
    \label{fig:complexity}
\end{figure}

In order to further detail this, Fig.~\ref{fig:complexity} compare the required complex multiplications per symbol detection for a different number of devices ($N$). Both of them show that AA-CL-DF has a lower complexity than other proposals with a better performance.

\begin{table}%
  \centering
  \subfloat[][Comparison of complexity order.]{\tabcomplex\label{tab:complexity}}%
  \qquad
  \subfloat[][Simulation setup.]{\tabvar\label{tab:var}}%
  \caption{Details of algorithms: (a) complexity and (b) simulation parameters.}%
  \label{tbl:table}%
\end{table}

\section{Simulation results}
\label{sec:sim_resul}

Averaging the result over $10^5$ runs, we consider an uplink
under-determined mMTC system with the setup parameters shown at
Table~\ref{tab:var}. All the simulated schemes experience an
independent and identically-distributed (i.i.d.) random flat-fading
channel model and the coefficients are taken from complex Gaussian
distribution of $\mathcal{C}\mathcal{N}\left(0,1\right)$.

\begin{figure}[t]
    \centering
        \includegraphics[scale=0.85]{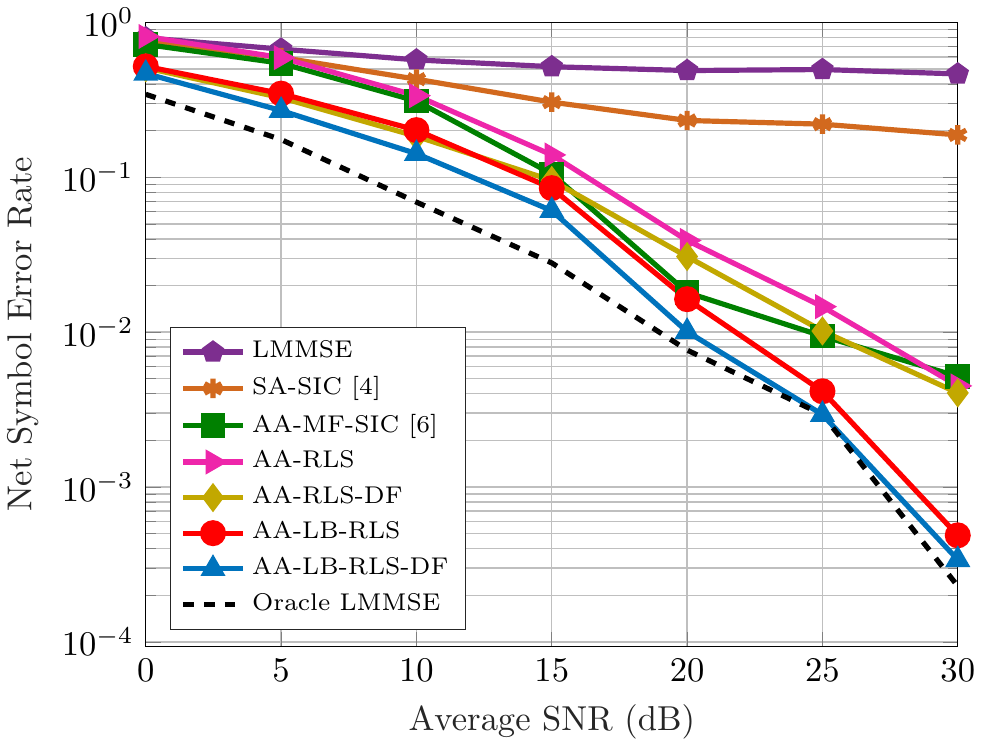}
    \caption{\footnotesize Symbol error rate vs. Average SNR. Perfect CSI, average SNR given by $10\log\left(N\, \sigma^2_x/\sigma^2_v\right)$, $\delta_\text{std}=0.5, \lambda_\text{std}=0.998$, $\delta_\text{reg}=0.7, \lambda_\text{reg}=0.92$, $\beta_\text{reg}=10$ and $\gamma=0.0001$.}
    \label{fig:result_wout_code_pf_csi}
\end{figure}

Fig.~\ref{fig:result_wout_code_pf_csi} shows the net symbol error
rate (NSER) which considers only the active devices. LMMSE exhibits
poor performance since the system is under-determined. Due to error
propagation, SA-SIC~\cite{Knoop2013} does not perform well. The
regularized version of RLS, AA-RLS presents a worse performance than
the AA-MF-SIC~\cite{DiRenna2019} but with the advantage of does not
need a explicit channel estimation. Inserting the modified
multi-feedback scheme in AA-RLS brings up a notable SER gain, as
seen in AA-CL-RLS results. The usage of the decision feedback
algorithm improves the performance of both proposals, in a way to
AA-CL-DF approaches the lower bound. We considered that the Oracle
LMMSE detector has the knowledge of the index of nonzero entries.

\begin{figure}[t]
    \centering
        \includegraphics[scale=0.85]{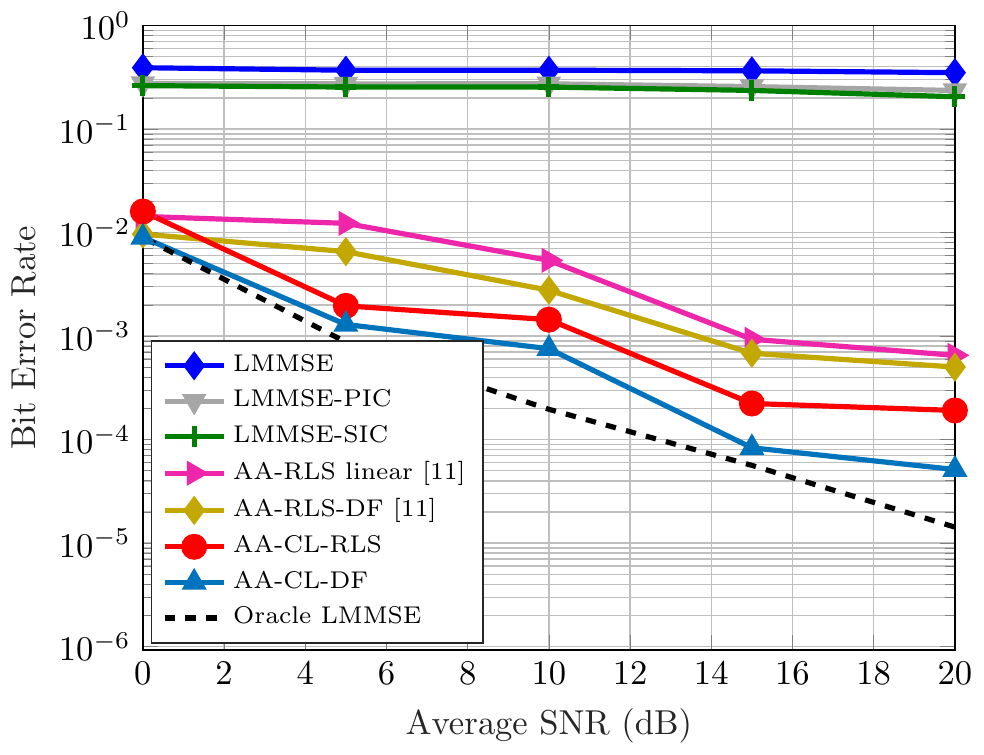}
    \caption{\footnotesize Bit error rate vs. Average SNR. Perfect CSI, average SNR given by $10\log\left(N\, \sigma^2_x/\sigma^2_v\right)$, $\delta_\text{std}=0.5, \lambda_\text{std}=0.998$, $\delta_\text{reg}=0.7, \lambda_\text{reg}=0.92$, $\beta_\text{reg}=10$ and $\gamma=0.0001$.}
    \label{fig:result_code_pf_csi}
\end{figure}

For the coded systems with IDD, the scheme of Wang and
Poor~\cite{XWang1999} is modified for mMTC as shown in
Section~\ref{sec:prop_it_det_dec}. We use a LDPC matrix with 256
columns and 128 rows, avoiding length-4 cycles and with 6 ones per
column \cite{memd}. The Sum-Product Algorithm (SPA) decoder is used
and the average SNR is $10\log\left(N R\,
\sigma^2_x/\sigma^2_v\right)$, where $R=1/2$ is the rate of the LDPC
code. Fig.~\ref{fig:result_code_pf_csi} compares the BER results of
the considered algorithms and the AA-CL-DF. The hierarchy of
performance of the other considered algorithms is the same as the
uncoded case but with better error rate values.

\begin{figure}[t]
    \centering
        \includegraphics[scale=0.85]{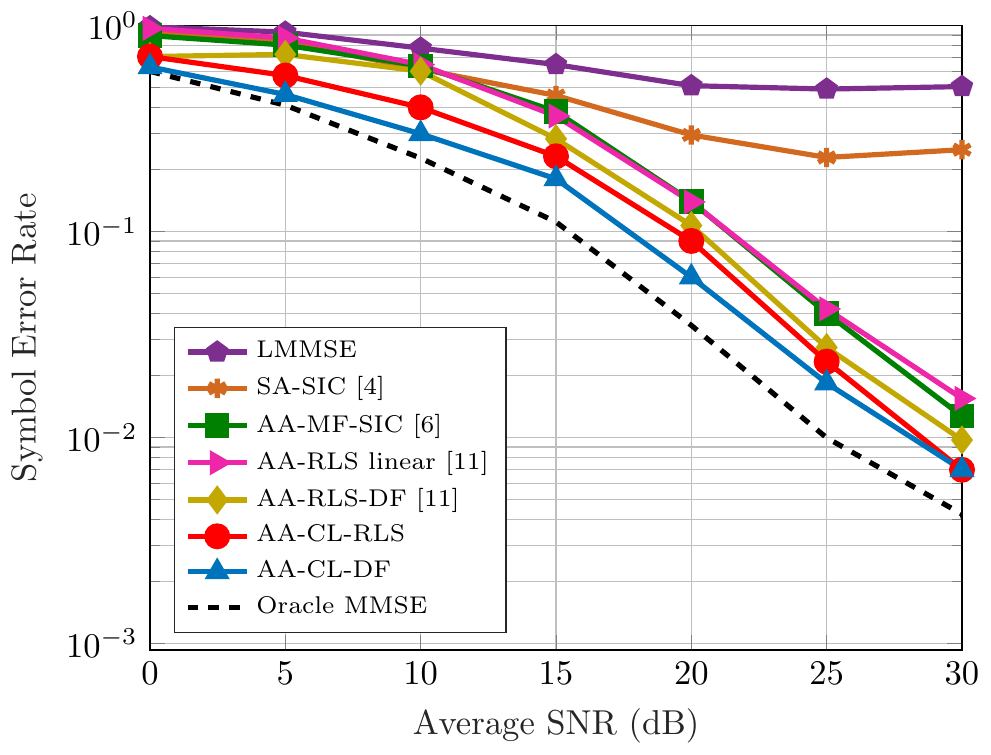}
    \caption{\footnotesize Symbol error rate vs. Average SNR. Imperfect CSI, average SNR given by $10\log\left(N\, \sigma^2_x/\sigma^2_v\right)$, $\delta_\text{std}=0.5, \lambda_\text{std}=0.998$, $\delta_\text{reg}=0.7, \lambda_\text{reg}=0.92$, $\beta_\text{reg}=10$ and $\gamma=0.0001$.}
    \label{fig:result_wout_code_impf_csi}
\end{figure}

\begin{figure}[t]
    \centering
        \includegraphics[scale=0.85]{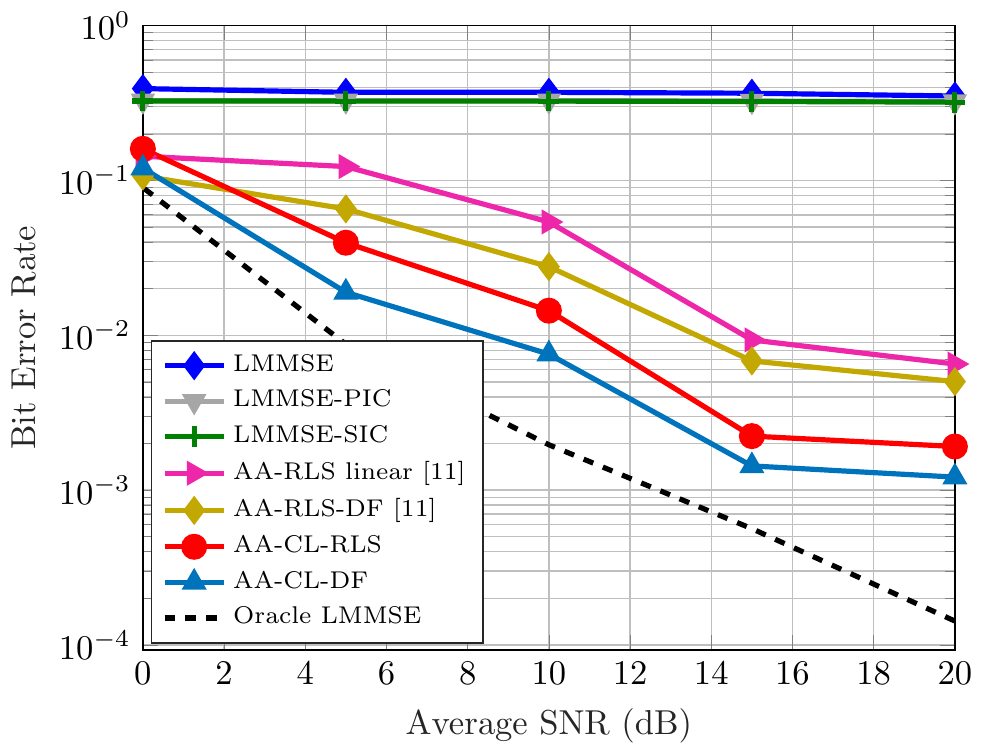}
    \caption{\footnotesize Bit error rate vs. Average SNR. Imperfect CSI, average SNR given by $10\log\left(N\, \sigma^2_x/\sigma^2_v\right)$, $\delta_\text{std}=0.5, \lambda_\text{std}=0.998$, $\delta_\text{reg}=0.7, \lambda_\text{reg}=0.92$, $\beta_\text{reg}=10$ and $\gamma=0.0001$.}
    \label{fig:result_code_impf_csi}
\end{figure}

For a comparison with imperfect channel estimation, we considered
$\hat{\mathbf{H}} = \mathbf{H} + \mathbf{E}$, where $\mathbf{H}$
represents the channel estimate and $\mathbf{E}$ is a random matrix
corresponding to the error for each link. Each coefficient of the
error matrix follows a Gaussian distribution, i.e., $\sim
\mathcal{C}\mathcal{N}\left(0,{\sigma}^2_e\right)$, where
$\sigma^2_e = \sigma_v^2/5$. The hierarchy of performance is almost
the same obtained in perfect channel estimation assumption,
differentiating only at the fact that the AA-MF-SIC is more suitable
to the imperfect CSI than the schemes with the regularized RLS, as
seen in Figs.~\ref{fig:result_wout_code_impf_csi}
and~\ref{fig:result_code_impf_csi}.

\section{Conclusions}\label{sec:Conc}

We have proposed the design of the AA-CL-DF scheme, for mMTC scenarios. In this context, compared with previous works, we have presented an $l_0$-norm regularized RLS decision feedback algorithm with the aid of a list based on constellation candidates. With a competitive complexity, simulations have shown that AA-CL-DF significantly outperforms existing approaches.


\end{document}